\documentclass[fleqn,usenatbib]{mnras}

\usepackage{newtxtext,newtxmath}

\usepackage[T1]{fontenc}

\DeclareRobustCommand{\VAN}[3]{#2}
\let\VANthebibliography\thebibliography
\def\thebibliography{\DeclareRobustCommand{\VAN}[3]{##3}\VANthebibliography}

\usepackage{graphicx}	
\usepackage{amsmath}	
\usepackage{float}
\usepackage{subcaption}
\usepackage{lipsum}

\graphicspath{{./Images}}

\title[Gravitational-wave millilensing]{Exploring the hidden Universe: A novel phenomenological approach for recovering arbitrary gravitational-wave millilensing configurations}

 \author[A. Liu et al.]{Anna Liu,$^{1}$\thanks{E-mail: ania.liu@link.cuhk.edu.hk (AL)}
Isaac C. F. Wong,$^{1}$
Samson H. W. Leong, $^{1}$
Anupreeta More, $^{2, 3}$
Otto A. Hannuksela,$^{1}$
\newauthor
and Tjonnie G. F. Li$^{1, 4, 5}$ \\
$^{1}$ Department of Physics, The Chinese University of Hong Kong, Shatin, NT, Hong Kong\\
$^{2}$ The Inter-University Centre for Astronomy and Astrophysics (IUCAA), Post Bag 4, Ganeshkhind, Pune 411007, India \\
$^{3}$ Kavli Institute for the Physics and Mathematics of the Universe (IPMU), 5-1-5 Kashiwanoha, Kashiwa-shi, Chiba 277-8583, Japan \\
$^{4}$ Institute for Theoretical Physics, KU Leuven, Celestijnenlaan 200D, B-3001 Leuven, Belgium\\
$^{5}$ Department of Electrical Engineering (ESAT), KU Leuven, Kasteelpark Arenberg 10, B-3001 Leuven, Belgium
 }

\date{Accepted XXX. Received YYY; in original form ZZZ}

\pubyear{2023}
\begin{document}
\label{firstpage}
\pagerange{\pageref{firstpage}--\pageref{lastpage}}
\maketitle

\begin{abstract}
\noindent
Since the first detection of gravitational waves in 2015, gravitational-wave astronomy has emerged as a rapidly advancing field that holds great potential for studying the cosmos, from probing the properties of black holes to testing the limits of our current understanding of gravity. One important aspect of gravitational-wave astronomy is the phenomenon of gravitational lensing, where massive intervening objects can bend and magnify gravitational waves, providing a unique way to probe the distribution of matter in the universe, as well as finding applications to fundamental physics, astrophysics, and cosmology. However, current models for gravitational-wave millilensing - a specific form of lensing where small-scale astrophysical objects can split a gravitational wave signal into multiple copies - are often limited to simple isolated lenses, which is not realistic for complex lensing scenarios. In this paper, we present a novel phenomenological approach to incorporate millilensing in data analysis in a model-independent fashion. Our approach enables the recovery of arbitrary lens configurations without the need for extensive computational lens modeling, making it a more accurate and computationally efficient tool for studying the distribution of matter in the universe using gravitational-wave signals. When gravitational-wave lensing observations become possible, our method could provide a powerful tool for studying complex lens configurations in the future. 
\end{abstract}




\begin{figure*}
\includegraphics[width=\textwidth]{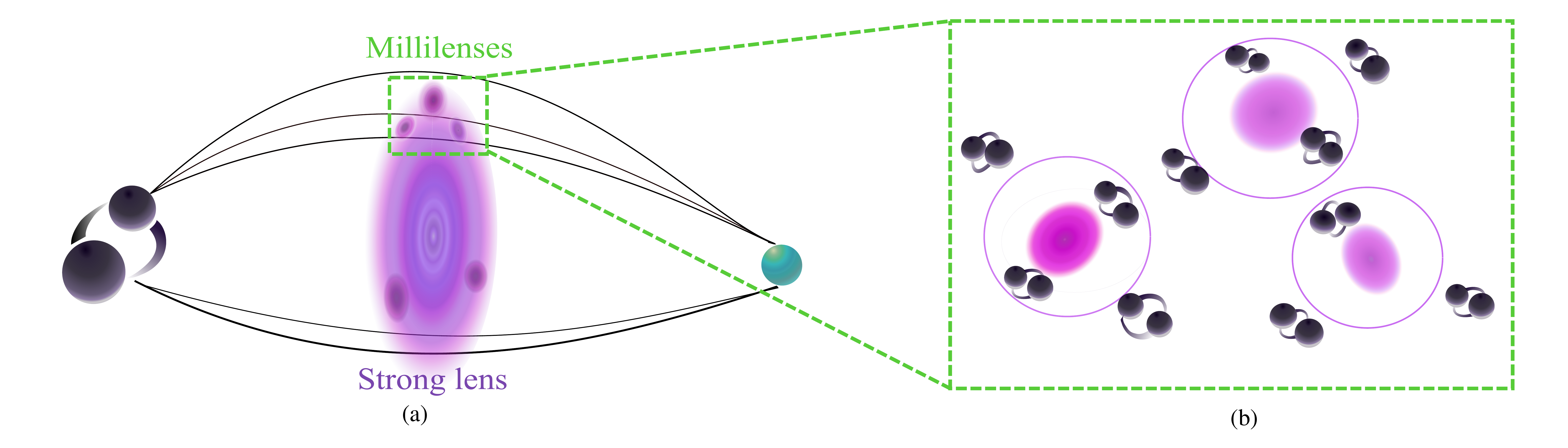}
    \caption{An example configuration of millilenses within a galaxy producing multiple millisignals and influencing each other gravitationally (figure not to scale); (a) as a gravitational wave passes near a massive object such as a galaxy or a galaxy cluster, its path gets bent which can result in multiple strongly-lensed GW signals (represented as curved black lines) reaching the observer; (b) additionally, if the signals encounter smaller compact objects, e.g., stars, dark matter subhalos, or massive compact halo objects, acting as millilenses, further splitting of the signals occurs, which results in multiple lensed BBH \textit{images} arising in the lens plane. The circular curves around each lens are the (tangential) critical curves. Throughout this work, we use the \textit{thin lens approximation} in which the source and lens objects are confined to two-dimensional planes.}
    \label{fig:set_up}
\end{figure*}

\section{Introduction}

Gravitational waves (GWs) were predicted by Albert Einstein's theory of general relativity in 1916. However, it took nearly a century before their direct detection by the LIGO and Virgo Collaborations in 2015, opening a new era of gravitational-wave astronomy \citep{abbott2016improved, abbott2016observation}. Since then, numerous GW detections have been made, including binary black hole (BBH) mergers, binary neutron star (BNS) mergers, and the merger of a BBH with a neutron star (BH-NS) \citep{abbott2019gwtc1, abbott2021gwtc2, abbott2021gwtc3} with the Advanced LIGO (\citet{aasi2015advanced}), Advanced Virgo (\citet{acernese2014advanced}), and KAGRA detectors (\citet{kagra2019kagra}).

Gravitational lensing, the bending of light or GWs by a massive object, was first observed in 1919 by Sir Arthur Eddington during a total solar eclipse \citep{Dyson:1920}, which confirmed Einstein's theory of general relativity. In recent years, the search for GW lensing signatures has become a fast-developing field. Multiple lensed signals can arrive at the observer, described as time-separated de/magnified GW signals with a phase shift relative to the unlensed GW signal \citep{Takahashi2003WaveBinaries, cao2014gravitational, Dai2017OnWaves}. The current gravitational-wave data has been used to search for GW lensing, but so far, there has not yet been any widely accepted detection.

The challenges in making GW lensing detections are significant, including the need for new tools to detect strong lensing~\citep{
Haris2018IdentifyingMergers,Dai2020SearchO2,Liu2020IdentifyingVirgo,Lo2021ASignals,Janquart2021AEvents,janquart2022golum}, micro- and millilensing~\citep{Lai2018DiscoveringLensing,Dai2018DetectingWaves,Liao2018AnomaliesSubstructures,Christian:3Gdetections,Pagano2020LensingGW:Waves,Kim2020IdentificationLearning,Seo2021StrongMicrolensing,Wright2021Gravelamps:Selection}, and wave optics lensing~\citep{Lai2018DiscoveringLensing, Christian:3Gdetections,Bulashenko2021LensingPatternb,Oguri2022AmplitudeLensingc,Basak2022ConstraintsMicrolensing,Tambalo2022LensingLenses}, beyond the need for statistical modelling of both gravitational lenses and binary black holes~\citep{Dai2016EffectMergers,Ng2017PreciseHoles,Smith2017WhatClusters,Smith2018Strong-lensingClusters,Oguri2018EffectMergers,Wierda2021BeyondLensing,Xu2021PleasePopulations,More2021ImprovedEvents,Smith2022DiscoveringObservatory}. Furthermore, the low expected rates of lensed GW events make detection difficult~\citep{ccalicskan2022lensing}. Nevertheless, recent studies have shown that the detection of lensed GW events is possible with current and future GW detectors, assuming the current best models for the strong lensing galaxy population and presuming that binary black holes trace the star-formation rate density \citep{Ng2017PreciseHoles, Li2018GravitationalPerspective, Oguri2018EffectMergers, Xu2021PleasePopulations, Wierda2021BeyondLensing,Wempe2022AObservations,Smith2022DiscoveringObservatory,Ma2022OnMergers}. Indeed, the era of gravitational wave lensing is imminent, and the detection of lensed GW events would provide valuable information about the properties of the lensing objects, test the theory of general relativity in a new regime, probe high-redshift cosmology, study dark matter, search for primordial black holes and MACHOs, precisely localize merging black holes, and more~\citep{Takahashi2005AmplitudeUniverse,Itoh2009AWaves,Baker2016Multi-MessengerWaves,collett2017testing,Liao2017PrecisionSignals,Fan2017SpeedSignals,Lai2018DiscoveringLensing,Dai2018DetectingWaves,Mukherjee2019Multi-messengerWaves,mukherjee2020probing,Diego2019ConstrainingFrequencies,Oguri2020ProbingWaves,Goyal2020TestingSignals,Hannuksela2020LocalizingLensing,Finke2021ProbingBinaries,Iacovelli2022ModifiedFunction,Chung2021LensingMass,Sereno2010StrongLISA,Bolejko2012Anti-lensing:Voids,hernandez2022measuring,Tambalo2022GravitationalMatter,Basak2022ConstraintsMicrolensing,Basak2022ProspectsHole} and the searches for gravitational-wave lensing have started recently~\cite{Hannuksela2019SearchEvents,LIGO_O3a,Li2019FindingEvents,Dai2020SearchO2,Liu2020IdentifyingVirgo,Pang2020LensedDetection,TheLIGOScientificCollaboration2021SearchRun,Kim2022Deep-2}. 

Depending on the lens mass, gravitational lenses can cause different types of lensing. Massive objects such as galaxies and galaxy clusters can produce strong gravitational lensing, resulting in multiple signals with varying time separations and effects on the GW signal~\citep{Takahashi2003WaveBinaries, Dai2017OnWaves, Smith2017WhatClusters, Smith2018DeepGW170814, Haris2018IdentifyingMergers, Liu2020IdentifyingVirgo, Robertson:2020, Ryczanowski:2020, Dai2020SearchO2, Wang2021IdentifyingDetectors, ezquiaga2021phase, Lo2021ASignals, Janquart2021AEvents, Janquart2021OnModes, Vijaykumar2022DetectionSignals, ccalicskan2022lensing, Cao2022DirectWaves}. On the other hand, smaller-mass lenses such as stars or stellar-mass compact objects can cause microlensing, resulting in potentially observable beating patterns in the frequency evolution of the GW waveform \citep{Deguchi:1986, Nakamura:1997sw, Takahashi2003WaveBinaries, Christian:3Gdetections, Jung:2019compactDM, diego2019observational, Mishra2021GravitationalGalaxies, meena2020gravitational, Cheung2020Stellar-massWaves, Bulashenko2021LensingPatternb, cremonese2021breaking, Seo2021StrongMicrolensing, Yeung2021MicrolensingMacroimages, Qiu2022AmplitudeWaves, Wright2021Gravelamps:Selection, Kim2022Deep-2}. In this paper, microlensing refers to lens masses of tens to hundreds of solar masses, while millilensing refers to lens masses in the range of $10^2-10^6 M_\odot$, covering dark matter subhalos, primordial black holes, massive compact halo objects (MACHOs), and lenses with an Einstein radius approximately the size of a milliarcsecond. The distinction between microlensing and millilensing is based on the lens mass range and the corresponding time delays between individual lensed GW signals. The analysis of microlensing requires the wave optics approximation in the low-mass regime, where the GW wavelength matches the Schwarzschild radius of the lens, while we assume millilensing follows the geometrical optics approximation in this work, which holds for the corresponding mass range~\citep{Takahashi2003WaveBinaries}.

\begin{figure}
    \centering
    \includegraphics[width=\columnwidth]{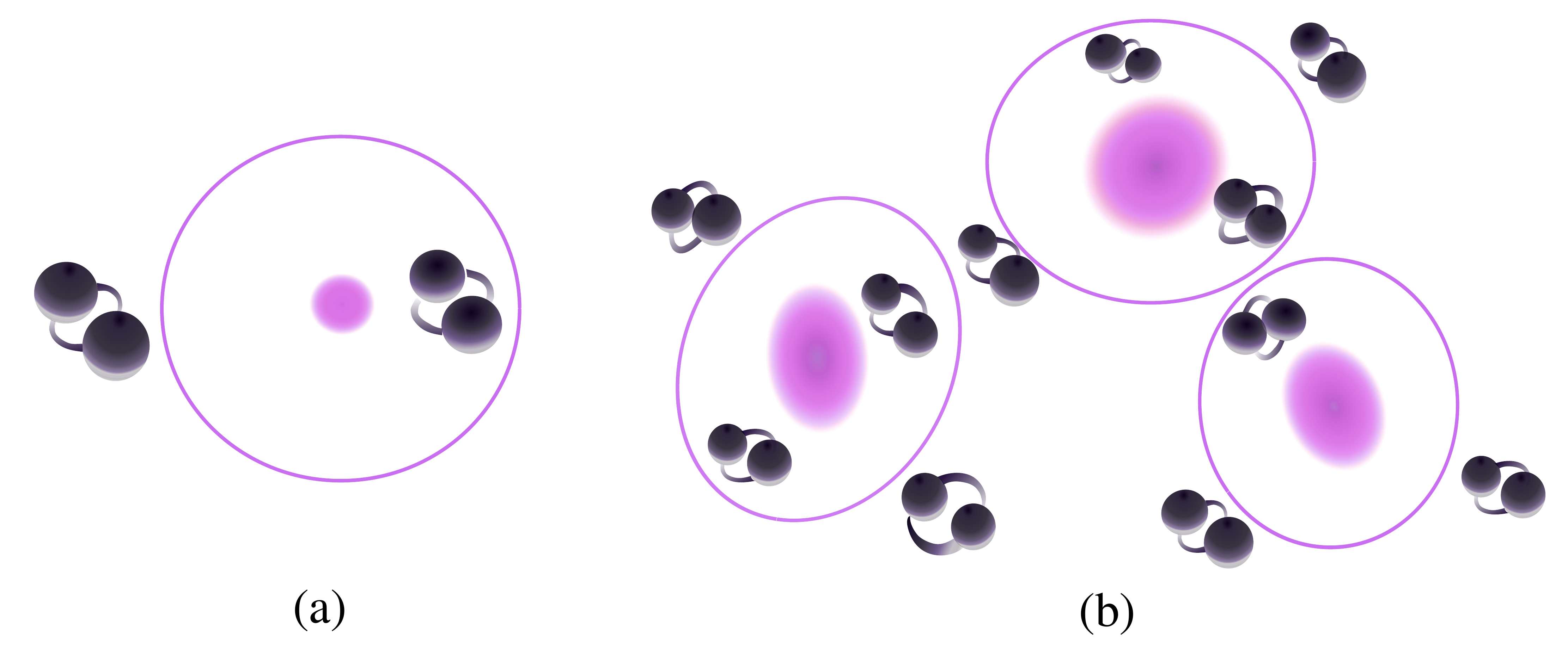}
    \caption{Schematic representation of point-mass lens model and phenomenological approach in the lens plane: (a) in the point-mass lens model, the lens is represented as a point-like mass (purple) producing two GW millisignals from the background BBH source. This model is often used in microlensing analysis, where the size of the lens is small compared to the relative size of the system and the critical curve of the lens is circular; (b) in the phenomenological approach presented here, we assume multiple millilenses can influence each other gravitationally, producing an overall gravitational shearing effect, hence breaking the spherical symmetry of the millilenses. Additionally, unlike the point-mass lens model, we assume each millilens can produce an arbitrary integer number of millisignals.
    }
    \label{fig:PMLvsPhenom}
\end{figure}

To date, several lens models have been suggested for GW micro- and millilensing studies. These models typically assume a particular lens mass distribution. Examples include the point-mass lens (PML) model and singular isothermal sphere (SIS) model (\citet{Nakamura:1997sw, Takahashi2003WaveBinaries, Keeton_book}), where the lenses are assumed to be isolated from any astrophysical objects. However, past electromagnetic lensing observations have shown that astrophysical objects, typically galaxies, can have a significant impact on the lens morphology, making it difficult to treat the lens as an isolated object~\cite[e.g.][]{diego2019observational,Seo2021StrongMicrolensing,Oguri2022AmplitudeLensingc}. For example, when a gravitational wave is strongly lensed by a galaxy and micro- or millilensed by the small-scale structure within the galaxy, the millilens would experience additional effects, including gravitational shear due to the galaxy's potential (see Fig.~\ref{fig:set_up})~\citep{Mishra2021GravitationalGalaxies,Seo2021StrongMicrolensing,Oguri2022AmplitudeLensingc}. This scenario could lead to a non-symmetric distribution of the millilens mass, as shown in Fig.~\ref{fig:PMLvsPhenom}. The widely used isolated spherically symmetric lens assumption is then physically unrealistic, at least in scenarios with significant shearing effects. In such a scenario, the current parameter estimation tools could obtain a significantly biased result or even miss the signal altogether~\citep{Yeung2021MicrolensingMacroimages}.

One potential way to improve our ability to recover these more complex lenses is to implement a collection of more intricate lens models in our parameter estimation framework, including extensions that account for galaxy shearing effects and fields of lenses, as well as several millilensing mass profiles. However, this approach would require a significant amount of effort to extend the parameter estimation framework to every plausible lensing scenario. Additionally, to implement this approach, we would need to perform parameter estimation on each signal for each lens model, making the process computationally expensive. 
Instead of relying on complex and computationally intensive lens models to account for physically realistic effects of astrophysical objects on millilensed gravitational-wave signals, we propose a different approach in this work. 

In particular, we suggest using a general, model-independent description of millilensed signals that parameterizes "image parameters" instead of the lensing system itself. Every millilensing system produces an integer number of gravitational-wave milli-images, each with their own set of magnifications, time delays, and arrival times, regardless of the lens configuration~\citep{Takahashi2003WaveBinaries,Dai2017OnWaves}. By parameterizing the properties of these milli-images, we can target any millilensing configuration, covering signals from different types of lenses and configurations without imposing limits on the system setup or the number of lensed GW signals formed. This phenomenological approach also allows us to recycle the results from the analysis of multiple data sets, without performing separate analyses with different lens models independently. Furthermore, we demonstrate that the results obtained from the phenomenological analysis parameterizing the lensed GW signal can be mapped to specific lens models to select the most favorable lens mass profile.

\section{Methods}
\label{sec:Methods}
\subsection{Lens models}
\label{sec:lensmodel} 
In order to demonstrate the effectiveness of our phenomenological model for searching generic millilensing configurations, we need to test it by injecting lensed GW signals into LVK detector data. To simulate these signals, we must choose a specific lens model, and for this exercise, we will focus on well-understood models that can be easily tested.

Therefore, we consider the point mass lens (PML) and singular isothermal sphere (SIS) models as our millilenses~\citep{Nakamura:1997sw, Takahashi2003WaveBinaries, Keeton_book}. We also incorporate an external shearing effect when these models are embedded into a galaxy. This effect stretches the caustic curves and can lead to additional milli-images~\citep{diego2019observational,Diego2019ConstrainingFrequencies}. While the PML model with a shearing effect is commonly used to approximate compact objects acting as gravitational lenses, the SIS model assumes a spherically symmetric mass distribution and is often used as a toy model to represent galaxies, star clusters, or dark matter subhalos. Note that more complex models, such as tidally stripped Navarro Fenk White lenses, may be required to accurately model dark matter subhalos, but our primary goal is to demonstrate a phenomenological search that can be applied to any lens model. We use the \textsc{Lenstronomy} package, a multi-purpose \textsc{Python} package for gravitational lensing, to perform all lens modeling~\citep{birrer2018lenstronomy}.

In addition to choosing a lens model, simulating the lensed GW waveform is also necessary. A lensed GW can be mathematically expressed as $\Tilde{h}_\textrm{L}(f)\!=\!F(f)\Tilde{h}(f)$, where $F(f)$ is the amplification factor, which is a function of frequency $f$ and specific lens parameters, and $\Tilde{h}(f)$ represents the unlensed GW. In the case of the PML and SIS models, the amplification factor is dependent on two lensed parameters: the redshifted lens mass $M{Lz}$ and the relative position of the source with respect to the lens, denoted by $y$. The addition of an external shearing effect, as described in the previous paragraph, would introduce three additional parameters: two components of the shearing effect, denoted as $\gamma_{1,2}$, and a convergence value denoted as $\kappa$. 

In the case of an isolated millilens, using the thin lens approximation and solving the lens equation for the PML model gives rise to two solutions of lensed GW signals (see the left panel in Fig.\ref{fig:PMLvsPhenom}). For relatively short time delays, the millilensed signals will overlap, leading to a single lensed GW signal arriving at the detector (see Fig.\ref{fig:waveform}). However, assuming the lens is embedded within a galaxy and including the internal shear, the number of the component millilensed GW signals can be different \citep{diego2019observational,Diego2019ConstrainingFrequencies}. Indeed, a generic lens potential may even include a field of small-scale lenses.

\subsection{Model-independent inference of millilensed GW signals}

\noindent
When considering millilensing, there are a multitude of potential lensing configurations, each with varying levels of complexity. This can include different types of millilenses, such as PMLs, SISs, or more complex lens structures, as well as a varying number of such lenses. Additionally, the effect of the strong lensing galaxy can further complicate the system. Due to this complexity, it is impractical to implement every gravitational lens model directly in parameter estimation. However, regardless of the specific millilens model, the observable signal will always consist of an integer number of GW millisignals, or "milli-images," each with their own properties such as time delay, magnification, and Morse phase. These signals overlap at the GW detector. In this work, we present a phenomenological search approach that directly targets the milli-images, allowing for an arbitrary configuration of the lensing system.

\begin{figure}
	\includegraphics[width=0.9\columnwidth]{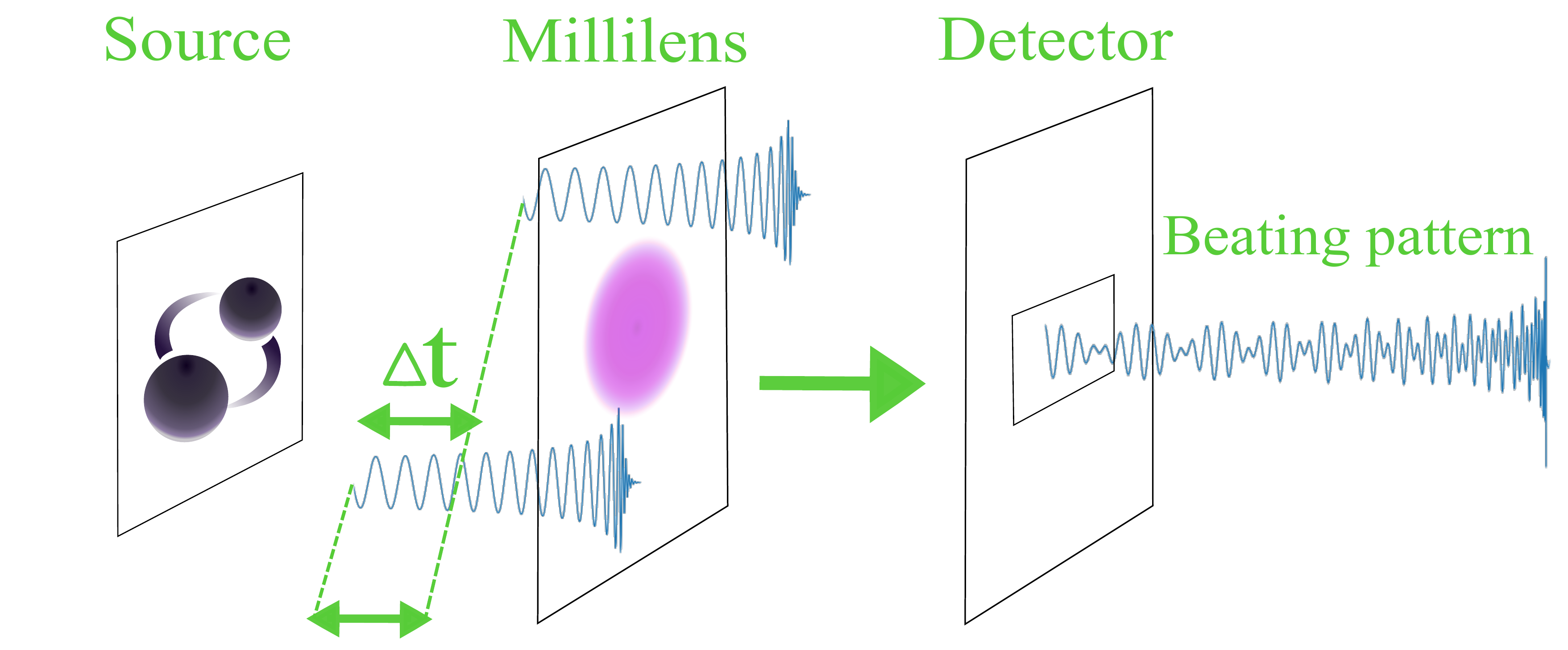}
    \caption{Illustration of the millilensing effect on the GW waveform. A GW signal passing near a millilens is split into two millisignals that overlap, resulting in a single GW signal at the detector with frequency-dependent beating patterns. One can discern the millilensing effect by locating the beating patterns in GWs. The thin lens approximation used confines the source, the lens and the observer to a 2-dimensional plane each. The deflection of the GW occurs instantaneously at the point where the GW crosses the lens plane.}
    \label{fig:waveform}
\end{figure}
 
When considering a millilens embedded in a macrosystem such as a galaxy, it is important to take into account the effects of the macrosystem. In contrast to previous gravitational wave analyses that have focused primarily on isolated point mass lenses and singular isothermal sphere profiles described by physical parameters of the lensing system (such as lens mass $M_{Lz}$ and source position $y$)~\cite[e.g.][]{TheLIGOScientificCollaboration2021SearchRun}, we parameterize each individual millilensed GW image using a set of millilensing parameters: relative magnification $\mu$, time delay $t$, and Morse factor $n$. This parameterization can account for non-symmetric effects due to the macrosystem without making prior assumptions about the system configuration and mass distribution. Each individual millilensed GW signal is treated independently of the other millisignals.

Before the individual millisignals reach the detector, they overlap due to relatively short time separations (order of milliseconds). The resultant GW signal can therefore be described as a sum of the millisignals with the waveform shape characterised by frequency-dependent beating patterns (see Fig.~\ref{fig:waveform}).

To allow for the most general case where any integer number of millisignals may be formed due to lensing, we introduce an additional lensing parameter, $K$, which corresponds to the number of millisignals. Unlike traditional models that assume a fixed number of millisignals based on a specific lensing scenario, our phenomenological approach does not limit the number of signals. By relaxing this assumption, we provide a more flexible framework for characterizing the complex physical configuration of the lensing system. The mathematical formulation of our approach is described below.

\subsection{A phenomenological formulation}
A gravitationally lensed waveform can be separated into an unlensed part (unlensed complex strain amplitude) multiplied by an \textit{amplification factor} $F(f, \;\boldsymbol{\theta_L})$ which contains lensing information described by lensing parameters $\boldsymbol{\theta_L}$. A millilensed waveform $\Tilde{h}_L(f; \boldsymbol{\theta}, \;\boldsymbol{\theta_L})$ can thus be expressed as
\noindent
\begin{align}
\begin{aligned}[c]
\tilde{h}_L(f; \boldsymbol{\theta}, \;\boldsymbol{\theta_L}) = {F(f, \boldsymbol{\theta_L})}\cdot \Tilde{h}_U(f; \boldsymbol{\theta}),
\end{aligned}
\end{align}
where  $\Tilde{h}_U(f; \boldsymbol{\theta})$ represents a frequency-domain GW waveform in the absence of millilensing with $\boldsymbol{\theta}$ corresponding to unlensed BBH source parameters, $F(f, \boldsymbol{\theta}_L)$ is an amplification function dependent on parameters associated with millilensing $\boldsymbol{\theta_L}=(\boldsymbol{\mu}, \boldsymbol{t}, \boldsymbol{n})$ and defined as:
\begin{equation}
\begin{aligned}[c]
    \label{eq:amplification_factor}
    F(f, \boldsymbol{\mu}, \boldsymbol{t}, \boldsymbol{n})= \sum_{j =1}^{K_\mathrm{max}}\left|\mu_{j}\right|^{1 / 2} \exp \left[2 \pi i f t_{j}-i \pi n_{j}\right],
\end{aligned}
\end{equation}
where the expression is summed over the total number of millisignals up to a chosen maximum number $K_\mathrm{max}$ and ($\mu_j$, $t_j$, $n_j$) correspond to the lensing parameters of the $j^{\mathrm{th}}$ millisignal. Such a summation corresponds to the \textit{geometrical optics approximation},  which applies when the GW wavelength $\lambda_\mathrm{GW}$ is longer than the Schwarzschild radius $R_S$ corresponding to the lens mass:  $\lambda_\mathrm{GW}>R_S$ (\citet{Deguchi:1986, Takahashi2003WaveBinaries}). For ground-based GW detectors, the geometrical optics approximation can be applied down to masses $M_{Lz}\sim \mathcal{O}(10^2)\,M_\odot$ (see Appendix~\ref{sec:geom_optics} for an explanation of the validity of geometrical optics approximation). Throughout this work, we use geometrical units ($c = G =1$).

The notation introduced in Eq.~\eqref{eq:amplification_factor} uses the conventional notation of lensing magnification $\mu$. However, to overcome degeneracy between consecutive millilensed signals, we use the effective luminosity distance notation which relates to the true luminosity distance from the source $d_L$ and the magnification of $j^{\mathrm{th}}$ lensed GW signal by:
\begin{equation}
    d^\mathrm{eff}_{j} = \frac{d_L}{\sqrt{\mu_{j}}}.
    \label{eq:eff_lum_dist}
\end{equation}
This effective luminosity distance notation allows for a better separation of individual millilensed signals and reduces degeneracy between the binary orbital and millilensing parameters. 

The individual millilensed GW signals must be time-ordered to avoid degeneracy between their arrival times. To achieve this, we define the time delay between each consecutive signal such that the arrival time of each signal increases with $j$, that is, $t_{j+1} > t_j$. We can describe the millilensed GW signals with respect to the first signal, which arrives earliest, by choosing $t_1=0$. Using the effective luminosity distance notation introduced in Eq.~\eqref{eq:eff_lum_dist}, we can express the amplification factor 
\begin{equation}
\begin{split}
    &F(f, d^\mathrm{eff}_j, t_j, n_j) \\ &= d_L\left(\frac{1}{d_1^\mathrm{eff}}\exp[-i\pi n_1] + \frac{1}{d^\mathrm{eff}_2}\exp\left[2\pi f t_2 - i\pi n_2\right]+\cdots \right)
\end{split}
\end{equation}
where the first signal (arriving the earliest) is described by its Morse factor $n_1$ and the consecutive millilensed signals ($j\ge 2$) are described by their effective luminosity distance $d^\mathrm{eff}_j$ (magnification) and time delay $t_j$ w.r.t the arrival time and the luminosity distance of the first signal.

Consequently, the millilensed GW signal can be expressed as 
\begin{equation}
\begin{split}
    \tilde{h}_{L}(f) &= \sum_{j =1}^{K_\mathrm{max}} \frac{d_L}{d^\mathrm{eff}_j} \exp \left[2 \pi i f t_{j}-i \pi n_{j}\right] \tilde{h}_{U}(f)(f; \boldsymbol{\theta}, d_L, t_\mathrm{col})\\
    &= \sum_{j =1}^{K_\mathrm{max}} \exp \left[ -i \pi n_{j}\right] \tilde{h}_{U}(f)^\mathrm{eff}(f; \boldsymbol{\theta}, d_L, t_\mathrm{col}, d^\mathrm{eff}_{\mathrm{j}}, t_{\mathrm{j}})
\end{split}
\end{equation}
where in the second line we have defined an effective GW signal dependent on the source parameters, as well as the millilensing parameters $({d^\mathrm{eff}_j}, {t}_j)$ of $j^\mathrm{th}$ component signal. 

The resultant millilensed GW signal, which is a sum of $K$ individual millilensed GW signals, can hence be expressed in terms of the amplification function as
\begin{equation}
     \tilde{h}_{L}(f; \boldsymbol{\theta}, \boldsymbol{\theta}^{L}_{K_\mathrm{max}}, K_\mathrm{max}) = F(f; \boldsymbol{\theta}^{L}_{K_\mathrm{max}}, K_\mathrm{max})\cdot \tilde{h}(f;\boldsymbol{\theta})
\end{equation}
where $\boldsymbol{\theta}^{L}_{K_\mathrm{max}}$ represents the lensing parameters of the sum of all $K_\mathrm{max}$ millilensed signals.

\subsection{Problem statement in Bayesian framework}
To perform parameter estimation and model selection on real GW data, we adopt a Bayesian framework. Our goal is to obtain a result with a specified number of millilensed GW signals, $K$, while considering varying dimensionality of the models due to the changing number of signals. To address this problem, we use joint Bayesian model selection, where the posterior distribution is defined on a union of subspaces with different dimensions, each corresponding to a model with a fixed number of signals. Assuming a countable set of $K_\mathrm{max}$ models with $k$ models, we define the total union of parameters as the BBH parameters $\boldsymbol{\theta}$ and lensing parameters $\boldsymbol{\theta}^{L}_{K_\mathrm{max}}$ with $K_\mathrm{max}$ being the maximum number of allowed signals specified by the user in each parameter estimation run. Our aim is to estimate the unknowns $k$ and $\boldsymbol{\theta}^{L}_k$, where $k\in{1, ..., K_\mathrm{max}}$, given the data set $\boldsymbol{d}$. We use the nested sampling algorithm (\citet{skilling2006nested}) to draw these unknowns from corresponding prior distributions.

We express the joint distribution of all variables as
\begin{equation}
    p(k, \boldsymbol{\theta}, \boldsymbol{\theta}^{L}_k, \boldsymbol{d}) = p(\boldsymbol{d}\vert \boldsymbol{\theta}, \boldsymbol{\theta}^{L}_k, k)p(\boldsymbol{\theta}) p(\boldsymbol{\theta}^{L}_k\vert k)p(k)
\end{equation}
where $p(\boldsymbol{d}\vert \boldsymbol{\theta}, \boldsymbol{\theta}^{L}_k, k)$ is the likelihood, $p(\boldsymbol{\theta}^{L}_k\vert k)$ is the prior distribution of lensing parameters, $p(\boldsymbol{\theta})$ is the prior distribution of the BBH source parameters and $p(k)$ is the prior distribution of the number of millilensed signals. 

We can express the marginal likelihood as follows (for detailed derivation, see Appendix (\ref{sec:appendix_derivations})):
\begin{equation}\label{eq:likelihood}
p(\boldsymbol{d})= \sum_{k=1}^{K_{\max }} \int p\left(\boldsymbol{d} \mid \boldsymbol{\theta}, \boldsymbol{\theta}^{L}_{k}, k\right) p(\boldsymbol{\theta}) p\left(\boldsymbol{\theta}^{L}_{k} \mid k\right) p(k) d \boldsymbol{\theta} d \boldsymbol{\theta}^{L}_{k}.
\end{equation}

We can obtain the joint posterior distribution $p(\boldsymbol{\theta}^{L}_k, k \vert \boldsymbol{d})$ and the posterior distribution for the number of images can be expressed as  
\begin{equation} \label{eq:posterior_k}
    p(k \mid \boldsymbol{d})=\frac{\int p\left(\boldsymbol{d} \mid \boldsymbol{\theta}, \boldsymbol{\theta}^{L}_{k}, k\right) p(\boldsymbol{\theta}) p(\boldsymbol{\theta}^{L}_{k} \mid k) p(k) d \boldsymbol{\theta} d \boldsymbol{\theta}^{L}_{k}}{\sum_{k=1}^{K_{\max }} \int p\left(\boldsymbol{d} \mid \boldsymbol{\theta}, \boldsymbol{\theta}^{L}_{k}, k\right) p(\boldsymbol{\theta}) p(\boldsymbol{\theta}^{L}_{k} \mid k) p(k) d \boldsymbol{\theta} d \boldsymbol{\theta}^{L}_{k}}.
\end{equation}

\section{Model-independent simulations}
\label{sec:sim}

\subsection{Simulations}

We begin by simulating a lensed GW waveform with varying numbers of millisignals using the \textsc{lenstronomy} package to model the physical setup of the system. The simulated signal is injected with BBH parameters from GW190408 and added to the detector Gaussian noise of the network of three ground-based detectors (LIGO Livingston, LIGO Hanford, and Virgo) using the \textsc{bilby} Bayesian inference library. We perform parameter estimation of both the source and millilensing parameters using the phenomenological waveform approximant \textsc{IMRPhenomPv2} and the nested sampling technique with the \textsc{dynesty} sampler. The priors used in the simulation are listed in Table~\ref{tab:priors}, with the time delay and magnification specified with continuous uniform priors and the Morse factor and number of images having discrete uniform priors. We simulate three injections with different signal-to-noise ratios (SNRs): 20, 30, and 50.

\begin{table}
	\centering
	\caption{Injected parameters for the SNR$= 20$ parameter estimation run. BBH source parameters' values correspond to the median values of GW190408 \citep{GWdata}, millilensing parameters are obtained from simulating a lensing system with \textsc{lenstronomy} package \citep{birrer2018lenstronomy, Birrer2021}} 
	\label{tab:injected_values}
	\begin{tabular}{ll}
		\hline
		BBH source parameter & Value \\ 
		\hline
		Mass 1, $m_1$ &  $31.6\,M_\odot$ \\
		Mass 2, $m_2$ &  $23.7\,M_\odot$ \\
            Luminosity distance $d_L$ & 1598 Mpc \\
		 Dimensionless Spin 1, $a_1$ & 0.35 \\
		 Dimensionless Spin 2, $a_2$ &  0.36 \\
          Tilt Angle 1, $\theta_1$ & 1.7 rad \\
          Tilt Angle 2, $\theta_2$ &  1.6 rad \\
          Right Ascension, $\mathrm{RA}$ & 6.088 rad \\
          Declination, $\delta$ & 0.92 rad \\
          Polarization, $\psi$ & 3. 16 rad\\
          Inclination, $\theta_\mathrm{jn}$ & 0.73 rad \\
          Azimuthal Angle of $\Vec{L}$, $\theta_\mathrm{jl}$ & 2.9 rad \\
          Azimuthal Angle Difference, $\phi_{12}$ & 3.1 rad \\
          Phase $\phi$ & 3.134 rad \\
		\hline
        Millilensing parameter & Value \\
        \hline 
        Effective luminosity distance $d^\mathrm{eff}_1$ & 1598 Mpc \\
        Effective luminosity distance $d^\mathrm{eff}_2$ & 1577 Mpc \\
        Effective luminosity distance $d^\mathrm{eff}_3$ & 2570 Mpc \\
        Effective luminosity distance $d^\mathrm{eff}_4$ & 2758 Mpc \\
        Time delay $t_2$ &  0.0066 s\\
        Time delay $t_3$ &  0.0467 s\\
        Time delay $t_4$ &  0.0512 s\\
        Morse phase $n_1$ & 0 \\
        Morse phase $n_2$ & 0\\
        Morse phase $n_3$ & 0.5\\
        Morse phase $n_4$ & 0.5 \\
        Number of millisignals $K$ & 4 \\
        \hline 
        
	\end{tabular}
\end{table}

\begin{table}
	\centering
	\caption{Prior distributions used for injection runs}
	\label{tab:priors}
	\begin{tabular}{ll}
		\hline
		Parameter & Prior distribution  \\ 
		\hline
		Luminosity distance $d_L$ & Uniform $\mathcal{U}$(50 Mpc, 20000 Mpc) \\
		Effective luminosity distance $d^\mathrm{eff}$ & Uniform $\mathcal{U}$(50 Mpc, 20000 Mpc) \\
		Time delay $dt$ & Uniform $\mathcal{U}$($10^{-3}$s, $10^{-1}$s) \\
		Morse factor $n$ & Discrete Uniform $\mathcal{U}\{0, 0.5, 1\}$ \\
		Number of millisignals $K$ & Discrete Uniform $\mathcal{U}\{1, 2, 3, 4, 5, 6\}$ \\
		\hline
	\end{tabular}
\end{table}

\subsection{Lens configuration}
We start with a strongly lensed GW signal by a galaxy resulting in two GW signals lensed at different deflection angles, and each of the two strongly lensed signals passes near a millilens located within the same plane as the strong lens, which experiences a gravitational shearing effect due to the presence of the strong lens galaxy. The modelling is divided into two steps: first, solving the lens equation for the strong lens to obtain the shear components ($\gamma_1, \gamma_2$), convergence components ($\kappa_1, \kappa_2$) and magnifications ($\mu_1^S, \mu_2^S$) of the two strongly lensed GW signals; secondly, using the shear values from a strong lens as a macro-environment input into millilens models to obtain the magnifications, time delays and Morse factors for the resultant millisignals which are then estimated in injection run parameter estimation. In this example, we assume that one of the strongly lensed signals is further split into four millilensed signals, but the choice of the number of millisignals is arbitrary and can take different integer values depending on the system setup, convergence, and shear values.

The strong lens is modelled as a galaxy with a singular isothermal ellipsoid (SIE) mass distribution located at redshift $z_\mathrm{lens} = 0.3$, with eccentricity components $e_1 = 0.05$ and $e_2 = 0$ and an Einstein radius of $\theta_E = 1$. The millilens is modelled as an SIS mass distribution, and the source is a BBH emitting GWs at typical redshift $z_\mathrm{source} = 1.5$ (\citet{Wierda:2021upe}). A flat $\Lambda$-CDM cosmological model is assumed throughout the simulation with $H_0 = 70 ; \mathrm{km s}^{-1} \mathrm{Mpc}^{-1}$ and $\Omega_\mathrm{M}=0.3$. Results of the injection run are presented in the following section.

\subsection{Example: 4 millisignals}

\begin{figure}
    \centering
    \includegraphics[width=\columnwidth]{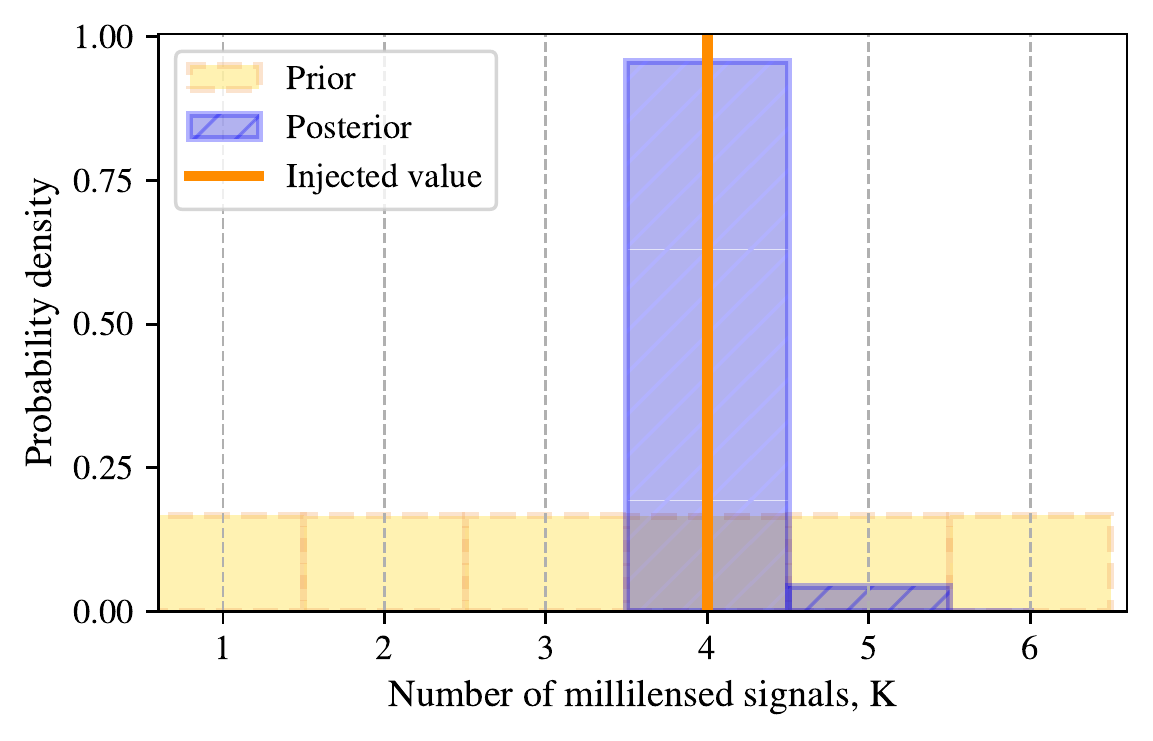}
    \caption{Number of millilensed signals, $K$, recovered from the injection run with uniform prior. The posterior distribution (shown in purple) is in agreement with the injected value $K=4$ (orange line). }
    \label{fig: no_of_images_K}
\end{figure}

The case reported here illustrates a GW signal lensed into 4 millisignals injected into detector noise. Fig.~\ref{fig: no_of_images_K} shows the posterior distribution for parameter $K$, corresponding to the number of millilensed GW signals recovered. All three runs with three different SNR values lead to consistent results, but for the clarity of the figure, only results from the injection run with SNR 20 are shown. As can be seen from the plot, the number of millisignals has been recovered in agreement with the injected value of $K=4$.

\begin{figure}
    \centering
    \begin{subfigure}{\columnwidth}
        \centering
        \includegraphics[width=\linewidth]{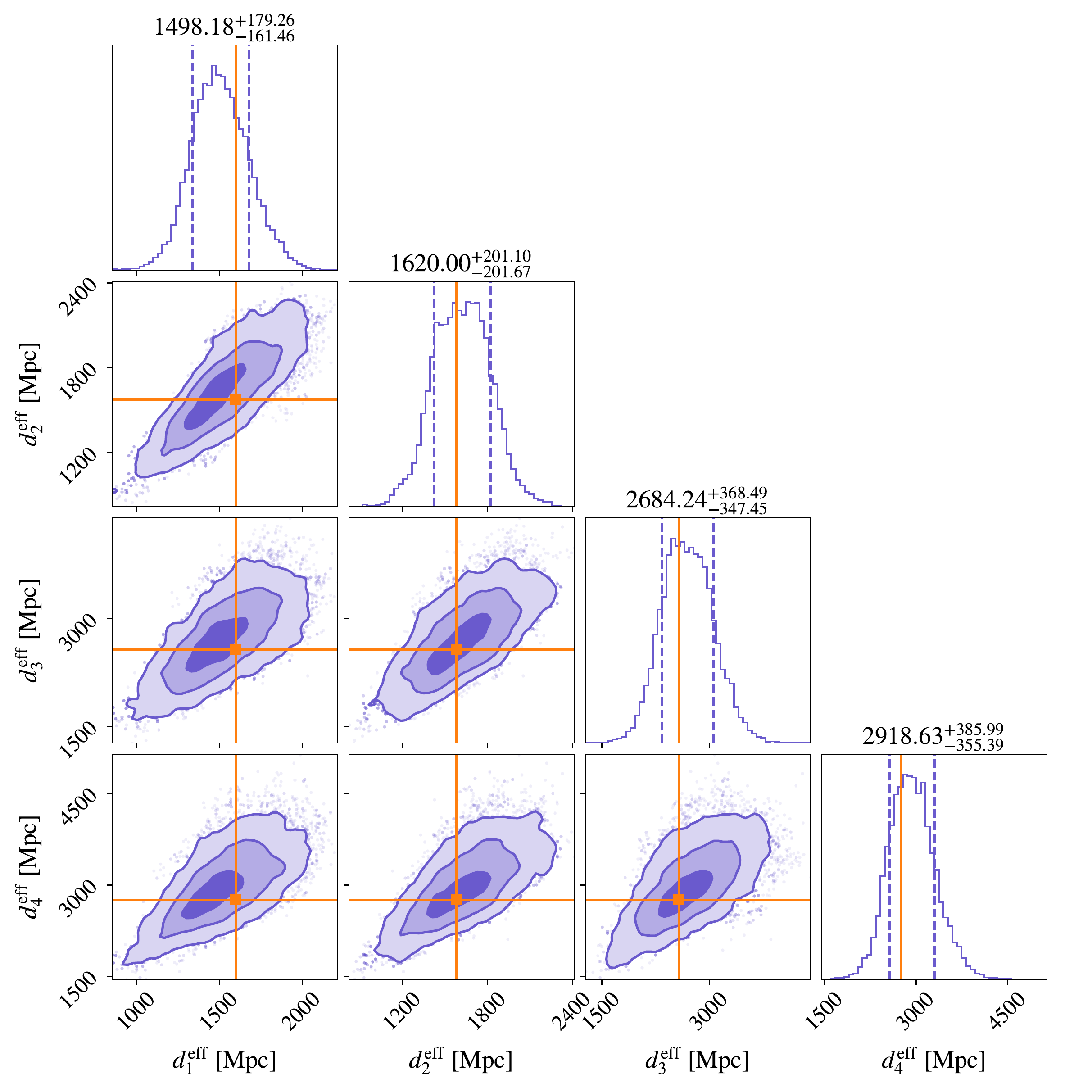}
        \caption{}
        \label{fig: eff_lum_distances}
    \end{subfigure}
    \par \bigskip
    \begin{subfigure}{\columnwidth}
        \centering
        \includegraphics[width=.8\linewidth]{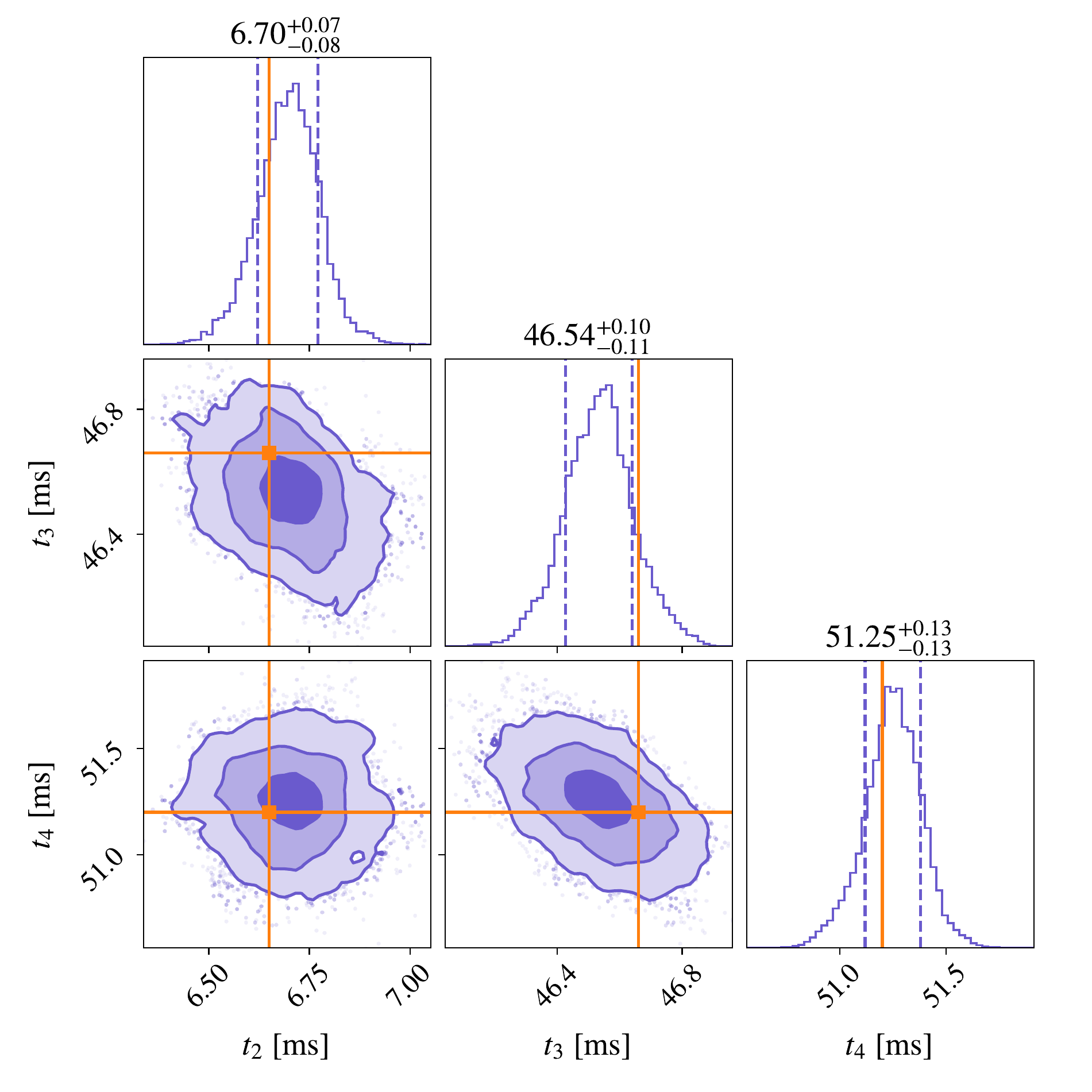}
        \caption{}
        \label{fig: time_delays}
    \end{subfigure}
    \caption{Results recovered from parameter estimation of the injected millilensed GW with an SNR 20 and GW190408 BBH parameters: (a) effective luminosity distances, (b) time delays of component millilensed signals w.r.t. the earliest arriving signal ($t_1 = 0)$. The millilens model is an SIS embedded into a strong lensing galaxy, which we model with the SIE lens model. The colours of the two-dimensional corner plots indicate 1$\sigma$, 2$\sigma$ and 3$\sigma$ credible regions. The orange lines show injected values for each parameter and the dashed lines in individual histogram plots correspond to 1$\sigma$ credible intervals. The effective luminosity distances and time delays are recovered with good accuracy and peak at the appropriate injected values.}
    \label{fig:my_label}
\end{figure}

\begin{figure}
    \centering
    \includegraphics[width=0.95\columnwidth]{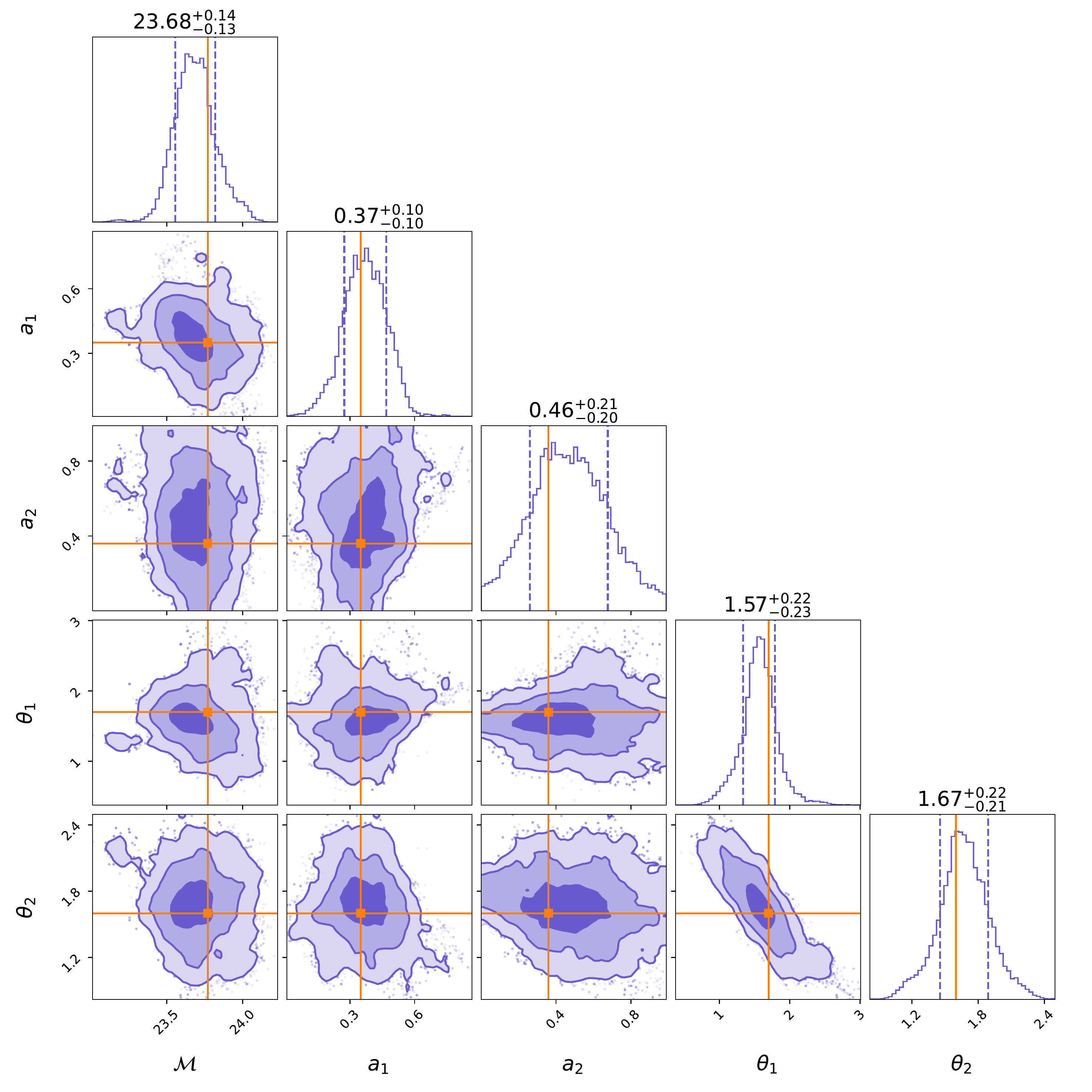}
    \caption{Black hole source parameters posterior distributions recovered from parameter estimation of the injected millilensed GW: chirp mass $\mathcal{M}$, dimensionless spin parameters $a_1$, $a_2$, tilt angles $\theta_1$, $\theta_2$. The orange lines represent the injected values for each parameter. The posterior distributions recovered are consistent with injected values.}
    \label{fig: BBHparams}
\end{figure}

Fig.~\ref{fig: eff_lum_distances} presents a corner plot of the effective luminosity distance parameters obtained for injection run with SNR 20. It can be seen from the figure that the injected values (orange lines) are well recovered in the posterior distributions for each parameter. Consistent results for higher SNR values were obtained.  
Similarly, the posterior distribution of time delays of the consecutive millilensed component signals are presented in Fig.~\ref{fig: time_delays}. The injected values are recovered within $1\sigma$ region of the posterior distributions for $t_2$ and $t_4$, and within $2\sigma$ for $t_3$. As tested with higher SNR runs, the accuracy of the recovered parameters increases with SNR. 
The recovered BBH source parameters are shown in Fig~\ref{fig: BBHparams}. 

\section{lens mapping}
\label{sec:mapping}

The methodology developed so far aims to provide a physically realistic picture of the lensing system, accounting for the fact that the lens located within a galaxy can be gravitationally affected by it. Having performed parameter estimation with the phenomenological approach, the results obtained can be mapped to specific lens models. We present an example mapping to an SIS model below. We choose the SIS model for its simplicity, which allows for analytical mapping from the phenomenological to SIS model parameters and vice versa. We use it as a testing example to demonstrate the accuracy of the method, noting that the method is not sensitive to physical assumptions and can be applied to more complex models. However, the drawback associated with the use of the SIS model is that it is a simplified model, which can not be applied to physically generic mass distributions if we want to take into account other physical effects, such as galactic shear. 

\subsection{Singular Isothermal Sphere}
In the geometrical optics limit, the time delay and magnification of two millilensed signals can be related to SIS lens parameters ($y$, $M_{Lz}$) by
\begin{equation}
\begin{aligned}
& t_{d}=8 M_{Lz} y\\
& \mu_{\pm}=\pm 1 + 1/y 
\label{eq:SIS}
\end{aligned}
\end{equation}
where $t_d$ is the time delay between two signals and $\mu_\pm$ are the magnifications of the two lensed signals which can be related to effective luminosity distances following Eq.~(\ref{eq:eff_lum_dist}), all expressed in $c=G=1$ units (\citet{Takahashi2003WaveBinaries}). 

The first step of the mapping is to extract the necessary data from the multi-signal millilensing analysis. If we consider $y\le 1$, the SIS model predicts that double-lensed GW signals are formed. Restricting ourselves to this case, we perform a millilensing injection run injecting two millilensed GW signals into detector noise, but allowing their number $K$ to take on values up to and including 6. From the analysis result, we select the posteriors of $(d^\mathrm{eff}_2, t_2)$ corresponding to $K=2$ case in order to map it to the predicted two-signal SIS model. Then, we construct the likelihood marginalized over all BBH parameters $\theta$ except for $(d^\mathrm{eff}_2, t_2)$:
\begin{equation}
    p(\boldsymbol{d}\vert d^{\mathrm{eff}}_2,t_2) = \int p(\boldsymbol{d}\vert d^{\mathrm{eff}}_2,t_2, \theta) p(\theta) d\theta.
\end{equation}
Knowing the relation between two sets of parameters (Eq.~(\ref{eq:SIS})), we assume that the likelihood can be expressed as
\begin{equation}
p(\boldsymbol{d}\vert d^{\mathrm{eff}}_2,t_2) =  p(\boldsymbol{d}\vert d^{\mathrm{eff}}_2 (y),t_2(M_{Lz}, y)) = p(\boldsymbol{d}\vert y, M_{Lz}).
\end{equation}
Then, using the likelihood, as well as assuming uniform prior distributions for ($y$, $M_{Lz}$), we perform a nested sampling over the two SIS parameters ($y$, $M_{Lz}$) in order to obtain corresponding posterior distributions and evidence for the SIS model. For the purpose of implementing the mapped likelihood into the Bayesian inference library \textsc{bilby}, we used the likelihood ratio to perform the nested sampling with the results presented in the following section. 

\begin{figure}
    \centering
    \includegraphics[width=\columnwidth]{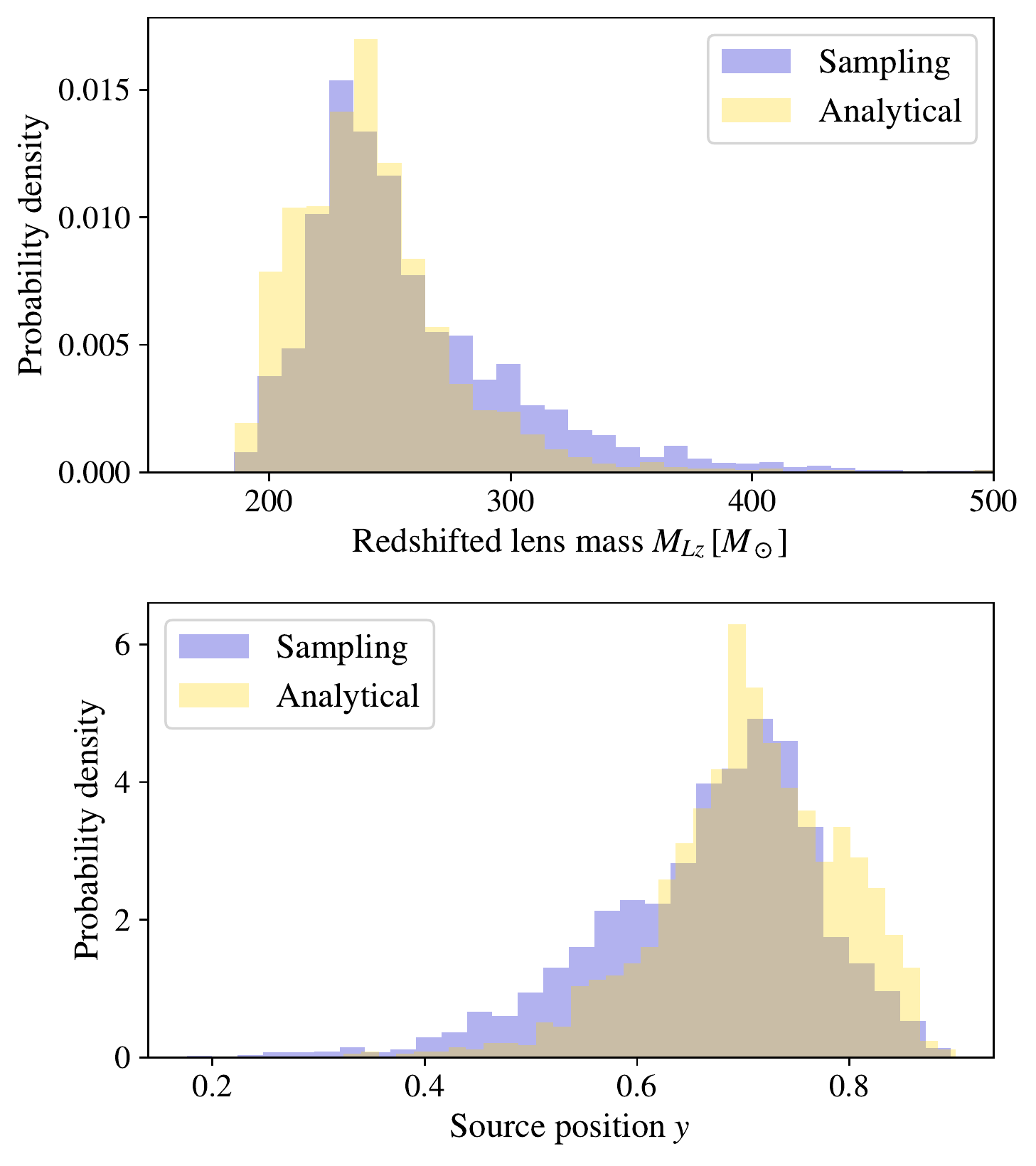}
    \caption{Posterior distribution of the redshifted lens mass $M_{Lz}$ (top) and  source position parameter $y$ (bottom), obtained from mapping phenomenological results to an SIS lens model. The mapping was performed: i) analytically (histogram in yellow), ii) by  nested sampling algorithm (histogram in purple). The results confirm that we are able to map the results from observables to lensing parameters with nested sampling.}
     \label{fig:SIS_plots}
\end{figure}

Following Eq.~(\ref{eq:SIS}), it is also possible to write down the inverse relations 
\begin{equation}
\begin{aligned}
&y = \frac{\mu_\mathrm{rel}-1}{\mu_\mathrm{rel}+1}\\
& M_{Lz} = \frac{t_{d}}{8} \frac{\mu_\mathrm{rel}+1}{\mu_\mathrm{rel}-1} 
\label{eq:SIS_inverse}
\end{aligned}
\end{equation}
where $\mu_\mathrm{rel}$ is the relative magnification $\mu_+/\mu_-$ between two lensed GW signals. For validation of the results obtained from nested sampling, we construct posterior distributions for the SIS parameters ($y, M_{Lz}$) by taking samples from the posterior distributions $p(t_d \vert \boldsymbol{d})$, $p(d^\mathrm{eff}_2 \vert \boldsymbol{d})$ and using the analytical relations Eq.~(\ref{eq:SIS_inverse}).

\subsection{Results}

Figure~\ref{fig:SIS_plots} represents the posterior distributions of SIS parameters $M_{Lz}$ and $y$, obtained from mapping phenomenological results. As described above, we used two methods to obtain the posterior distributions of the SIS parameters. Firstly, we performed a nested sampling algorithm (plots shown in purple in Fig.~\ref{fig:SIS_plots}). Secondly, to validate the results, we also used analytical relations from Eq.~(\ref{eq:SIS_inverse}) to construct the posteriors for $M_{Lz}$ and $y$ (plots in yellow in Fig.~\ref{fig:SIS_plots}). The prior ranges of ($M_{Lz}, y$) used in the two methods were different, hence the shape and position of the peak of the two distributions are not exactly the same (Fig.~\ref{fig:SIS_plots}). Nevertheless, the distributions obtained with the two methods lead to consistent results.

\section{Discussion and conclusion}
\label{sec:Conclusions}

\subsection{Discussion}
In this work, we have presented a new phenomenological approach to GW millilensing analysis, applicable to astrophysical lenses with masses in the range $10^3 - 10^6\, M_\odot$. The novel approach, unlike currently most widely used models, parameterizes the lensed GW signals, which not only enables to include the physical effects due to the presence of other massive objects in the vicinity of the gravitational lens but also provides an efficient lens model selection tool. Having tested the feasibility of the approach we applied the method to simulated injections of millilensed GW signals into detector noise. The parameter estimation of the lensed BBH parameters recovered results that are in agreement with injected values. 

The question raised by this study aims to provide a physically realistic description of an astrophysical lens within a galaxy. However, the generalizability of the results is subject to the limitation of assuming a single-lens system. More broadly, the study should be repeated addressing systems with multiple lenses located in the vicinity of each other (within the lens plane) or located along the line of sight. This would be a fruitful area for further work, providing a more generic description of gravitationally millilensed systems in different physical configurations.

Moreover, further research should be undertaken to investigate the distinction of millilensing effects on the gravitational waveform from other physical processes, such as spin precession or non-GR effects, which may also lead to frequency-dependent beating patterns in the waveform, mimicking gravitational lensing. It would be of crucial importance to distinguish between those effects once a GW signal with a potentially lensed waveform is detected. 

The methodology can be applied to a range of problems within theoretical physics, astrophysics and cosmology. A potential interesting example application is the problem of dark matter subhalos. 
Despite many observations supporting the presence of dark matter, the nature of dark matter remains one of the key open questions within the field (\citet{zackrisson2010gravitational, ellis2010lensing, Bertone2016AMatter, bertone2018new}). To date, a number of dark matter models have been proposed and some of them predict dark matter halos to be formed hierarchically from smaller subhalos which can further be formed from even smaller subhalos 
(\citet{moore1999dark, Metcalf2001CompoundGL, Liao2018AnomaliesSubstructures, Dai2018DetectingWaves}). However, current dark matter models show discrepancies at the smallest scales, often referred to as the substructure crisis (\citet{Metcalf2001CompoundGL, somerville2002can, moore2006globular}). In particular, some models predict dark matter subhalos cannot be present below a certain scale (\citet{kravtsov2010dark, Oguri2020ProbingWaves}). Therefore, in order to test the most feasible dark matter models, it is necessary to study dark matter subhalos down to the smallest scales. Moreover, if dark matter subhalos with masses of order $10^3 - 10^6\; M_\odot$ exist within galaxies as predicted by some models, they could potentially act as millilenses, directly influencing distant GW signals and producing lensing distortions. The probe of dark matter subhalos with gravitational millilensing has been proposed within EM studies of lensed quasars \citep{Wambsganss1992, Mao1997EvidenceGalaxies, Metcalf2001CompoundGL}, however, due to telescope resolution and propagation effects, EM observations can be subject to uncertainties. GW millilensing is subject to different systematics and could potentially be used as a direct measurement of the unknown dark matter subhalo mass function. Therefore, the dark matter substructures problem is a promising direction to study with the millilensing approach developed, which could become an alternative probe of the nature of dark matter at small scales. However, more investigation is needed to study the detailed capabilities of GW millilensing as a probe of dark matter subhalos.

\subsection{Conclusions}

The present study was designed to develop a phenomenological approach to gravitational millilensing studies which accounts for effects not included in currently used lens models. The simulations confirmed the feasibility of the method and the results support the idea that it is possible to study millilens configurations embedded in macro systems with mutual gravitational interactions. These findings suggest that in general, we can analyse non-isolated and non-symmetric gravitational lenses. Furthermore, results obtained from millilensing analysis can be mapped to existing lens models, providing a useful tool to distinguish between the most feasible models. These results add to the rapidly expanding field of gravitational-wave lensing and will prove useful in developing analysis tools for observational GW data with lensed GW detections predicted to take place in the coming years. The major limitation of this study is the geometrical optics approximation which limits the millilens mass range considered and the target lens population. Notwithstanding the relatively limited lens sample, this work offers valuable insights into the studies of gravitational millilensing analysis and can be further developed into population studies of potential millilens candidates. Non-observation of millilensing of predicted lenses at typical redshifts could also shed light on lens populations. The approach can also be expanded into generic millilensing studies of multiply lensed GW signals. Moreover, the millilensing framework could be applied to existing problems in astrophysics and cosmology, such as studies of the dark matter subhalos and primordial black holes with masses expected to lie within the corresponding millilensing mass range. More work will be needed, however, to study the detailed science case.

\section*{Acknowledgements}
The work is partially supported by grants from the Research Grants Council of the Hong Kong (CUHK 14306218), The Research Foundation -- Flanders (G086722N) and KU Leuven (STG/21/061). The analysed data and the corresponding power spectral densities are publicly available at the online Gravitational-Wave Open Science Center~\citep{GWOSC}. The authors are grateful for computational resources provided by the CIT cluster of the LIGO Laboratory and supported by National Science Foundation Grants PHY-0757058 and PHY-0823459. This material is based upon work supported by NSF's LIGO Laboratory which is a major facility fully funded by the National Science Foundation. This manuscript has LIGO DCC number P2200365.

\section*{Data Availability}
This research has made use of data or software obtained from the Gravitational Wave Open Science Center (\url{gwosc.org}), a service of LIGO Laboratory, the LIGO Scientific Collaboration, the Virgo Collaboration, and KAGRA. LIGO Laboratory and Advanced LIGO are funded by the United States National Science Foundation (NSF) as well as the Science and Technology Facilities Council (STFC) of the United Kingdom, the Max-Planck-Society (MPS), and the State of Niedersachsen/Germany for support of the construction of Advanced LIGO and construction and operation of the GEO600 detector. Additional support for Advanced LIGO was provided by the Australian Research Council. Virgo is funded, through the European Gravitational Observatory (EGO), by the French Centre National de Recherche Scientifique (CNRS), the Italian Istituto Nazionale di Fisica Nucleare (INFN) and the Dutch Nikhef, with contributions by institutions from Belgium, Germany, Greece, Hungary, Ireland, Japan, Monaco, Poland, Portugal, Spain. KAGRA is supported by Ministry of Education, Culture, Sports, Science and Technology (MEXT), Japan Society for the Promotion of Science (JSPS) in Japan; National Research Foundation (NRF) and Ministry of Science and ICT (MSIT) in Korea; Academia Sinica (AS) and National Science and Technology Council (NSTC) in Taiwan.




\bibliographystyle{mnras}
\bibliography{references}




\newpage
\onecolumn
\appendix
\center

\section{Validity of geometrical optics approximation}
\label{sec:geom_optics}

Throughout this work, we have assumed the geometrical optics approximation. In order to verify the validity of the approximation in the corresponding mass range considered, we compute the match between the lensed GW in geometric and wave optics for a PML with redshifted lens mass $M_{Lz}\in [10, 10^3]\,M_\odot$ and source position $y\in[0.01, 1]$. We define the match as


\begin{equation}
     \mathcal{M}(\Tilde{h}^L_\mathrm{wave}, \Tilde{h}^L_\mathrm{geo}) := \frac{\left(\Tilde{h}^L_\mathrm{wave}, \Tilde{h}^L_\mathrm{geo}\right)}{\sqrt{(\Tilde{h}^L_\mathrm{wave}, \Tilde{h}^L_\mathrm{wave})(\Tilde{h}^L_\mathrm{geo}, \Tilde{h}^L_\mathrm{geo})}}
\end{equation}
where the numerator is the noise-weighted inner product between the waveforms in two approximations 
\begin{equation}
   \left(\Tilde{h}^L_\mathrm{wave}, \Tilde{h}^L_\mathrm{geo}\right)= 4\;\Re \int_{f_\mathrm{low}}^{f_{\mathrm{high}}} \frac{\Tilde{h}^L_\mathrm{wave}(f) \Tilde{h}^{\ast L}_\mathrm{geo}(f)}{S_n(f)} df
\end{equation}

with $S_n(f)$ corresponding to the power spectral density (PSD) of the noise (in this case we use aLIGO design sensitivity PSD with frequency range [10, 2048] Hz), and the lensed GWs described as $\Tilde{h}^L_\mathrm{wave}(f) = F_\mathrm{wave}(f)\Tilde{h}(f)$  for wave optics effects and $\Tilde{h}^L_\mathrm{geo}(f) = F_\mathrm{geo}(f)\Tilde{h}(f)$ in geometrical optics approximation where the amplification functions are described respectively by 

\begin{equation}
F_\mathrm{wave}(f) = \exp  {\left[\frac{\pi w}{4}+i \frac{w}{2}\left(\ln \left(\frac{w}{2}\right)-2 \phi_{m}(y)\right)\right] }\times \Gamma\left(1-\frac{i}{2} w\right){ }_{1} F_{1}\left(\frac{i}{2} w, 1 ; \frac{i}{2} w y^{2}\right)
\label{eq:complete}
\end{equation}
where $w=8 \pi M_{L z} f ; \phi_m(y)=\left(x_m-y\right)^2 / 2-\ln x_m$ with $x_m=\left(y+\sqrt{y^2+4}\right) / 2 ; M_{L z}=$ $M_L\left(1+z_L\right)$ is the redshifted lens mass and ${ }_1 F_1$ is the confluent hypergeometric function (\citet{Takahashi2003WaveBinaries}); in the geometrical optics limit, the amplification function is 
\begin{equation}
F_\mathrm{geo}(f)=\left|\mu_{+}\right|^{1 / 2}-i\left|\mu_{-}\right|^{1 / 2} e^{2 \pi i f \Delta t_{d}}
\label{eq:geo}
\end{equation}
where the magnifications of the two lensed images are 
$\mu_{\pm}=1 / 2 \pm\left(y^{2}+2\right) /\left(2 y \sqrt{y^{2}+4}\right)$ and the time delay between two lensed signals is 
\begin{equation*}
    \Delta t_{d}=4 M_{L z}\left[y \sqrt{y^{2}+4} / 2+\ln \left(\left(\sqrt{y^{2}+4}+y\right) /\left(\sqrt{y^{2}+4}-y\right)\right)\right].
\end{equation*}

\begin{figure}
    \centering
    \includegraphics[width=0.54\textwidth]{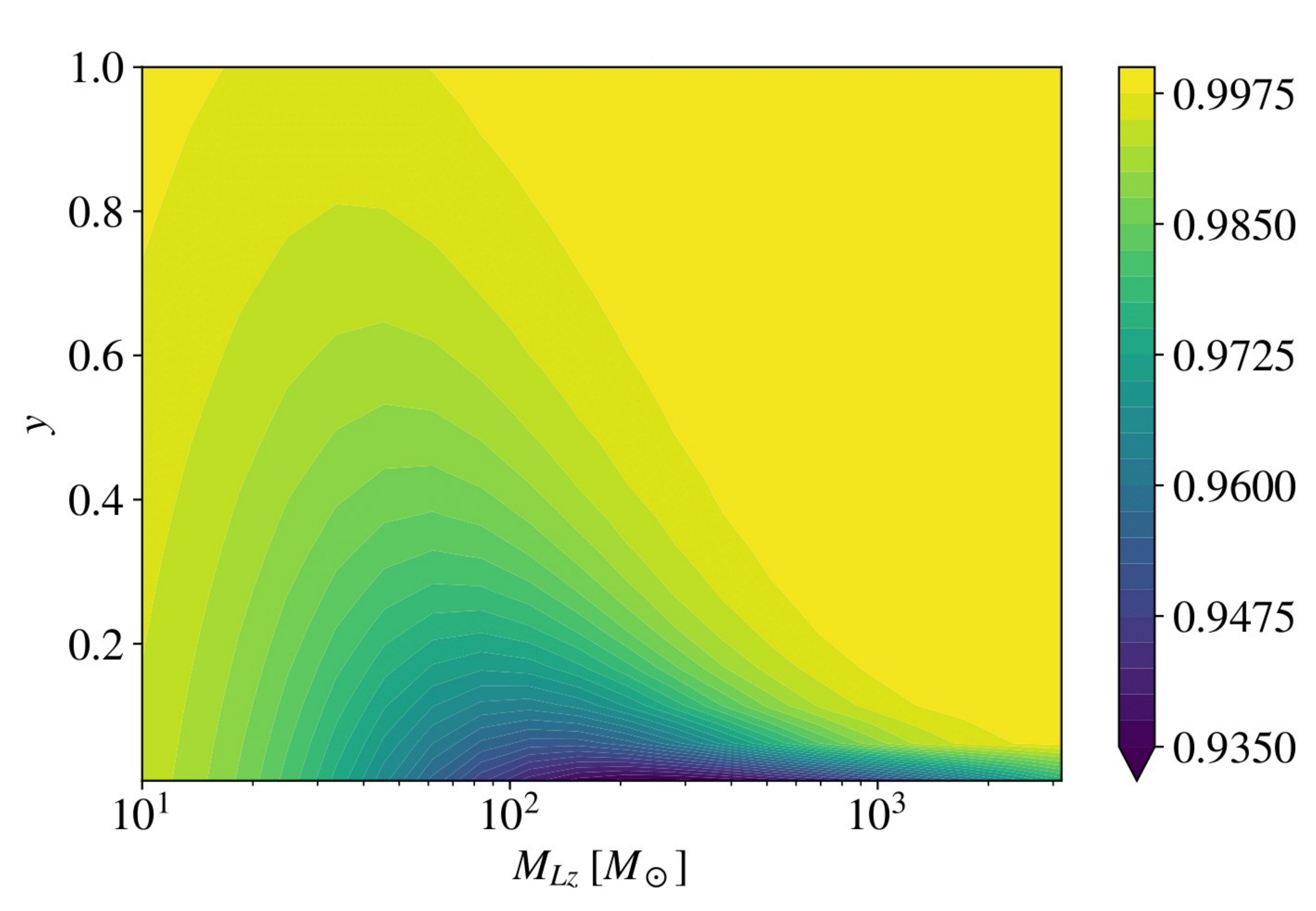}
    \caption{Contour plot representing the match between the plus polarisation of a lensed GW in geometrical optics approximation and the one of a lensed GW accounting for wave optics effects, plotted as a function of the relative source position $y$ and redshifted lens mass $M_\mathrm{Lz}$. The lens is assumed to be a PML.}
    \label{fig:geo_optics}
\end{figure}

From the match between the two waveforms, we can conclude that the geometric and wave optics effects match over 95\% for lens masses of order 200 $M_\odot$ (see Fig.~\ref{fig:geo_optics}). Therefore, we can conclude that geometrical optics approximation is valid for the mass range considered in the millilensing analysis presented in this paper: $M\in[10^3, 10^6] M_\odot$.

\section{Derivations}
\label{sec:appendix_derivations}
The following section presents derivation for the marginal likelihood of the millilensing methodology. 

The marginal likelihood of the millilensing model with an unknown number of images takes the general form
\begin{equation}\label{eq:1}
\begin{split}
p(\boldsymbol{d})&= \sum_{k=1}^{K_{\max }} \int p\left(\boldsymbol{d} \mid \boldsymbol{\theta}, \boldsymbol{\theta}^L_{k}, k\right) p(\boldsymbol{\theta}) p\left(\boldsymbol{\theta}^L_{k} \mid k\right) p\left(\boldsymbol{\theta}^L_{-k} \mid k\right) p(k) \,d \boldsymbol{\theta} d \boldsymbol{\theta}^L_{k} d \boldsymbol{\theta}^L_{-k}\\
&=\sum_{k=1}^{K_{\max }} \int p\left(\boldsymbol{d} \mid \boldsymbol{\theta}, \boldsymbol{\theta}^L_{k}, k\right) p(\boldsymbol{\theta}) p\left(\boldsymbol{\theta}^L_{k}, \boldsymbol{\theta}^L_{-k} \mid k\right) p(k) \,d \boldsymbol{\theta} d \boldsymbol{\theta}^L_{k} d \boldsymbol{\theta}^L_{-k} \\
&=\sum_{k=1}^{K_{\max }} \int p\left(\boldsymbol{d} \mid \boldsymbol{\theta}, \boldsymbol{\theta}^L_{k}, k\right) p(\boldsymbol{\theta}) p(\boldsymbol{\theta}^L_k \mid k) p(k)\, d \boldsymbol{\theta} d \boldsymbol{\theta}^L_k
\end{split}
\end{equation}
where $\boldsymbol{\theta}^L_{k}$ corresponds to the lensing parameters of $k$ millilensed GW signals, $\boldsymbol{\theta}^L_{-k}$ represent millilensing parameters of those millilensed signals for which the actual number of millilensed signals formed is greater than $k$. The two terms are assumed to be independent of each other in the prior distribution. We provide an illustrative example in the following section. 
\section*{Two-signal example}
Let us consider an example with $K_{\mathrm{max}} = 2$. Following equation (\ref{eq:1}), we can write
\begin{equation}
    \begin{split}
    p(\boldsymbol{d}) &= \sum_{k=1,2} \int p\left(\boldsymbol{d} \mid \boldsymbol{\theta}, \theta_1^L, \theta_2^L, k\right) p(\boldsymbol{\theta}) p(\theta_1^L)  p(\theta_2^L) p(k) d \boldsymbol{\theta} d\theta_1^L d\theta_2^L\\
    &= \int p\left(\boldsymbol{d} \mid \boldsymbol{\theta}, \theta_1^L, \theta_2^L, k=1\right) p(\boldsymbol{\theta}) p(\theta_1^L)  p(\theta_2^L) p(k=1) d \boldsymbol{\theta} d\theta_1^L d\theta_2^L + \int p\left(\boldsymbol{d} \mid \boldsymbol{\theta}, \theta_1^L, \theta_2^L, k=2\right) p(\boldsymbol{\theta}) p(\theta_1^L)  p(\theta_2^L) p(k=2) d \boldsymbol{\theta} d\theta_1^L d\theta_2^L\\
    &= \int p\left(\boldsymbol{d} \mid \boldsymbol{\theta}, \theta_1^L, k=1\right) p(\boldsymbol{\theta}) p(\theta_1^L)  p(k=1) d \boldsymbol{\theta} d\theta_1^L \textcolor{blue}{\int p(\theta_2^L) d\theta_2^L} + \int p\left(\boldsymbol{d} \mid \boldsymbol{\theta}, \theta_1^L, \theta_2^L, k=2\right) p(\boldsymbol{\theta}) p(\theta_1^L)  p(\theta_2^L) p(k=2) d \boldsymbol{\theta} d\theta_1^L d\theta_2^L
\end{split}
\end{equation}

where in the second line we have written out explicitly that each of the terms corresponding to a different value of $k$, two terms in this case for $k=1$ and $k=2$. Furthermore, the integral corresponding to $k=1$ term is split into two parameter spaces: $\theta_1$ and $\theta_2$, where the integral over $\theta_2$ indicated in blue gives 1 by definition. 
Hence, the conditional marginal likelihood of $k = 1$ is not affected by the lensing parameters for $k > 1$.


\bsp	
\label{lastpage}
\end{document}